\documentclass[aps,prl,amsmath,floatfix,twocolumn]{revtex4}
\usepackage{fontawesome}
\usepackage{amsmath, bm}
\usepackage{graphicx}
\usepackage{dcolumn}
\usepackage{dirtytalk}
\usepackage{color}
\usepackage{amsmath}
\usepackage{here}
\usepackage[utf8x]{inputenc}
\usepackage{bm}
\usepackage[mathscr]{euscript}
\usepackage{amssymb}
\usepackage{amsmath}

\usepackage{hyperref}
\usepackage{titlesec}
\usepackage{comment}
\usepackage{dsfont}
\usepackage{slashbox}
\usepackage{array}
\usepackage{bm}
\usepackage{braket}
\usepackage{MnSymbol}
\usepackage{graphicx}
\usepackage{pifont}
\usepackage[scr=rsfs]{mathalpha}

\hypersetup{breaklinks=true,
            colorlinks=false,
            pdfborder={0 0 0}}
\graphicspath{{figures/}}

\newcolumntype{L}[1]{>{\raggedright\let\newline\\\arraybackslash\hspace{0pt}}m{#1}}
\newcolumntype{C}[1]{>{\centering\let\newline\\\arraybackslash\hspace{0pt}}m{#1}}
\newcolumntype{R}[1]{>{\raggedleft\let\newline\\\arraybackslash\hspace{0pt}}m{#1}}
\newcommand{\cmark}{\ding{51}}
\newcommand{\xmark}{\ding{55}}

\renewcommand{\vec}[1]{\mathbf{#1}}

\begin{document}

%\title{Layer photovoltaic effect and intrinsic out-of-plane photocurrent in layered heterostructures}

\title{Layer photovoltaic effect in van der Waals heterostructures}

\author{Oles Matsyshyn, Ying Xiong, Arpit Arora, Justin C. W. Song}
\email{justinsong@ntu.edu.sg}

\affiliation{Division of Physics and Applied Physics, School of Physical and Mathematical Sciences, Nanyang Technological University, Singapore 637371,
}

\begin{abstract}
We argue that the layer electric polarization of noncentrosymmetric layered heterostructures can be generically controlled by light yielding a layer photovoltaic effect (LPE). The LPE possesses a rich phenomenology and can arise from myriad distinct mechanisms displaying strong sensitivity to symmetry (e.g., point group and time-reversal) as well as the presence/absence of a Fermi surface. We systematically classify these and unveil how LPE manifests for a range of light polarizations and even for unpolarized light. %Surprisingly, although electrons in layered heterostructures are confined to move within the two-dimensional plane without an out-of-plane velocity, they can nevertheless host an intrinsic out-of-plane photocurrent. 
These unusual layer photoresponses can be realized in a range of layered heterostructures such as bilayer graphene aligned on hexagonal Boron Nitride and manifest sizeable layer polarization susceptibilities in the terahertz frequency range that can be used as novel means of bulk photodetection.
\end{abstract}

\maketitle
Mechanical stacks of atomically thin van der Waals (vdW) materials enable to build quantum phases from the bottom-up with properties that go beyond that of its individual constituent components~\cite{balents2020superconductivity, Song2018electron}.  A particularly striking example is the emergence of a layer degree of freedom in stacks. Manipulating the  relative degree with which each of the layers is charged, as characterized by its static interlayer polarization, affords means to dramatically engineer bandstructure~\cite{zhang2009direct,tong2017topological}, 
tune quantum geometric properties~\cite{yao2008valley, song2016giant,yin2022tunable, ma2022intelligent}, 
as well as realize corrrelated phases of matter~\cite{chen2019evidence, zhou2021half, zhou2021superconductivity,de2022cascade}. 
Since {interlayer polarization} points out-of-plane, it is highly sensitive to vertical displacement fields. As a result, it has been traditionally controlled by toggling voltages sustained across a dual top and bottom gate sandwich architecture~\cite{zhang2009direct}. 
 
\begin{figure}[t]
    \centering
    \includegraphics[width=0.47\textwidth]{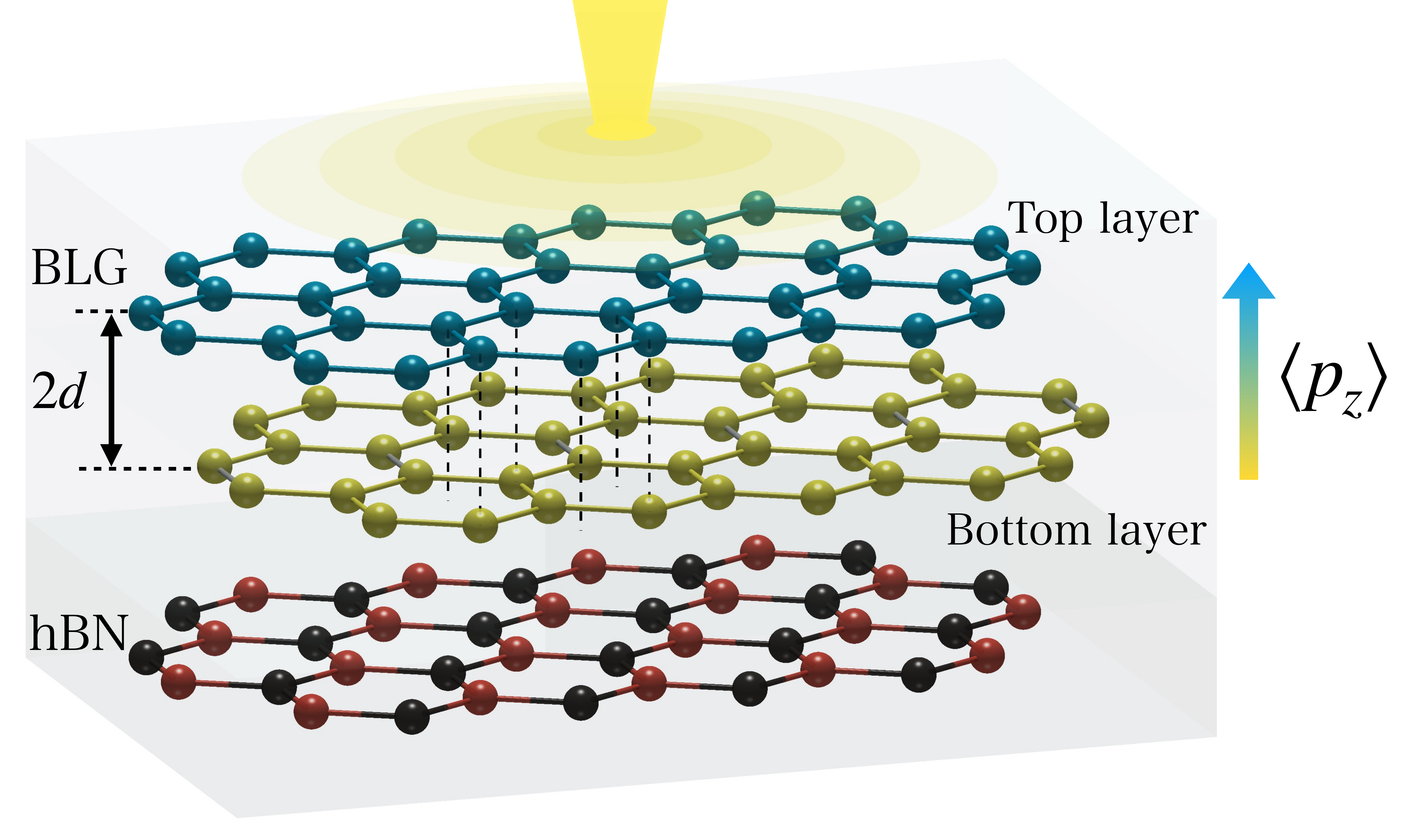}
    \caption{{\it Layer photovoltaic effect and interlayer polarization.} Photoinduced nonlinear interlayer polarization (here denoted by $p_z$) in noncentrosymmetric van der Waals stacks; we term this the layer photovoltaic effect (LPE). Here an example of a non-centrosymmetric and achiral vdW structure is shown: bilayer graphene aligned with hexagonal Boron Nitride (BLG/hBN). These achiral structures possess LPE induced by non-helical light. }
    \label{fig1}
\end{figure}

Here we argue that interlayer polarization in noncentrosymmetric layered heterostructures can be generically controlled by light manifesting a {\it layer} photovoltaic effect (LPE). Such LPE responses appear second-order in the incident light electromagnetic (EM) field, and, as we show below, 
come in myriad distinct types; by performing a systematic classification we delineate LPEs with distinct 
symmetry constraints, light polarization dependence, as well as physical origins. Importantly, we find LPEs can arise from both resonant interband absorption as well as off-resonant virtual processes in either metallic or insulating states, providing versatile means to control interlayer polarization across different phases of matter. 

We note that an example of LPE was recently predicted in chiral vdW bilayers where the interband absorption of circularly polarized light in such handed stacks induces a helicity dependent photoinduced {interlayer polarization}~\cite{PhysRevLett.124.077401}. Our work systematically shows that there exist a 
wide range of LPE interlayer responses beyond those known previously. 
For instance, at high frequencies corresponding to interband transitions we find an injection-like process enables {\it non-helical} light to induce a (second-order) nonlinear LPE even in an 
{\it achiral} and noncentrosymmetric vdW layered heterostructure, see Fig.~\ref{fig1}. Surprisingly, the injection-like process also produces an interlayer current even as the electrons do not possess an out-of-plane velocity. Instead, this intrinsic interlayer current arises from the pumping of the interlayer polarization. Additionally, even at low frequency without interband transitions, we find new types of large LPE responses that can be induced in the metallic regime. As we will see, these latter metallic contributions arise from the momentum-space asymmetry in the layer polarization of Bloch states on the Fermi surface. 

We anticipate that the LPEs we unveil can be used in novel bulk photodetection schemes that do not require p-n junctions. Since many non-centrosymmetric vdW stacks are achiral possessing mirror symmetries that render helicity dependent LPE vanishing, non-helical LPEs are crucial in activating interlayer polarization responses. Indeed, as we discuss below, the injection and metallic LPEs we unveil in our work can achieve giant susceptibility values in bilayer graphene (BLG) aligned with hexagonal boron nitride (hBN) heterostructures, orders of magnitude larger than those reported in chiral stacks~\cite{PhysRevLett.124.077401} and manifesting even for unpolarized light.

\textit{\color{blue}Interlayer polarization response}. We begin by directly examining the LPE response which is directly connected to the layer degree of freedom, $l$. 
The {interlayer polarization} operator is 
\begin{equation}
    \hat{P}^z= e d\sum_{\alpha_\ell}\hat{l}\ket{\alpha_l}\bra{\alpha_l}=\hat p^z d ,
    \label{eq:LPEoperator}
\end{equation}
where $\hat{l}$ is the 
layer index operator: $\hat l \ket{\alpha_l} = l\ket{\alpha_l}$, and $\ket{\alpha_l}$ are orbitals localized on layer $l$. For clarity, here we concentrate on a bilayer system with an interlayer distance $2d$ (see Fig.~\ref{fig1}).
Our theory, however, is general and can be readily applied to multi-layered systems. 

When light is normally incident on the vdW stack (see Fig.~\ref{fig1}), an out-of-plane static {interlayer polarization} can be induced. To see this, first consider the Hamiltonian: $\hat{H}(\mathbf{k},t) = H_{0}(\mathbf{k})+H_E(\mathbf{k},t)$ where $H_{0}(\vec k)$ is the bare Hamiltonian with $|u_{n{\bf k}}\rangle$ and $\epsilon_n(\vec k)$ the corresponding Bloch states and eigenenergies; here and below, roman indices denote band indices. $H_E(\mathbf{k},t)$ describes the light-matter interaction. For a monochromatic EM field, $H_E(\mathbf{k},t) = e\hat{\mathbf{r}}\cdot\left[\vec E e^{i\Omega t} + \vec{E}^* e^{-i\Omega t}\right]e^{\eta t}$~\cite{PhysRevB.52.14636,PhysRevB.61.5337}, with $\hat{\mathbf{r}}$ the position operator, $\eta\rightarrow0^+$ an adiabatic turn-on parameter, and $\Omega$ is the frequency of the light. 

The LPE can be obtained from 
Eq.~(\ref{eq:LPEoperator}) as $\langle P^z(t)\rangle=\int {\rm{Tr}}[\hat\rho(t) \hat P_z]d\mathbf{k}/(2\pi)^2$ where $\hat\rho$ is the density matrix. Here the evolution of the density matrix and the resulting {photoinduced interlayer polarization} can be tracked in a standard perturbative fashion, see full details in Supplementary Information ({\bf SI}). This produces a second-order nonlinear photoinduced static interlayer polarization, $\langle \delta {P}^z _{\rm st}\rangle$, characterized by an LPE susceptibility tensor $\chi (\omega)$ as: 
\begin{equation}
\langle \delta {P}^z _{\rm st}\rangle = 2d\sum_{\alpha\beta}{\rm Re}\left[E^\alpha E^{\beta*}\chi^{\alpha\beta}(\Omega)\right],
\label{eq:LPEfreq}
\end{equation}
where $\alpha,\beta$ are spatial indices ($x$ or $y$). 
We will show there are five contributions to $\chi^{\alpha\beta}(\omega)$ with distinct physical origins, symmetry properties, and phenomenology. 

To proceed, it is useful to delineate between interband and intraband responses and for concreteness we will confine ourselves to band non-degenerate systems.
Three contributions comprise interband responses: Injection (I), Shift (S) and Fermi-Sea (FS): 
\begin{equation}
    \chi_{\rm{inter}}^{\alpha\beta}(\omega) = \chi_{\rm{I}}^{\alpha\beta}(\omega)+\chi_{\rm{S}}^{\alpha\beta}(\omega)+\chi_{\rm{FS}}^{\alpha\beta}(\omega),
\end{equation}
where $\chi_{\rm I}$ and $\chi_{\rm S}$ describe LPE arising from resonant real interband excitations, whereas $\chi_{\rm FS}$ is off-resonant. 

The injection susceptibility $\chi_{\rm{I}}^{\alpha\beta}(\omega) = \tau \sigma_{\rm inter}^{\alpha\beta} (\omega)/4$ with 
\begin{equation}
    \sigma_{\rm inter}^{\alpha \beta} (\omega) = \frac{\pi e^2}{\hbar^2} 
    \sum_{n,m, {\vec k}}\delta(\omega+\omega_{nm})
    A^\alpha_{nm}A^\beta_{mn}f_{nm} \delta P_{mn},
    \label{eq:inj}
\end{equation}
where $f_{nm}=f[\epsilon_n(\vec k)]-f[\epsilon_m(\vec k)]$ is the difference between Fermi functions in different bands, $\hbar \omega_{nm} = \epsilon_n (\vec k) -\epsilon_m (\vec k)$, and $\delta P_{mn} = p^z_{mm} -p^z_{nn}$ is the difference between layer polarization between the final and initial states states. $p^z_{nm} = \langle u_{n{\bf k}}|\hat p^z |u_{m{\bf k}}\rangle$ is a matrix element of the polarization operator and $A_{nm}^\alpha=i \langle u_{n{\bf k}}|\partial_{\bf k_\alpha} |u_{m{\bf k}}\rangle$ is the interband 
Berry connection~\cite{Blount}. Here $\tau$ is a phenomenological relaxation time~\cite{PhysRevX.10.041041} that regularizes the $\chi_{\rm I}$ response.

\begin{table}[t]
\centering
\resizebox{0.45\textwidth}{!}{
\begin{tabular}{|C{0.8cm}||C{0.8cm}|C{0.8cm}| C{0.8cm} | C{1.4cm} | C{1.7cm} |}
\hline\hline{}&{}&{}&{}&{}&{}\\
  {}& \text{SC}  &\text{FS}&\text{shift }&{\text{injection}}&\text{reported in} \\ 
 {}&{}&{}&{}&{}&{}\\ \hline\hline{}&{}&{}&{}&{}&{}\\
 \faArrowsV & \cmark  &\cmark & \xmark & \cmark &\text{this work}\\
 {}&{}&{}&{}&{}&{}\\\hline{}&{}&{}&{}&{}&{}\\
 \faRepeat & \xmark  & \xmark & \cmark & \xmark &Ref.\cite{PhysRevLett.124.077401}\\
 {}&{}&{}&{}&{}&{}\\\hline
\end{tabular}
}
\caption{\label{tab:Table1} {\it LPE mechanisms in TRS preserving systems}. \faArrowsV~ indicates non-helical mechanisms (induced by linearly polarized light), while ~\faRepeat~ indicates helical responses (induced by circularly polarized light). \cmark~ denotes allowed, \xmark~ indicates forbidden. SC and FS are semiclassical and Fermi Sea respectively. Note that $\chi_{\rm B}$ (see text) is forbidden when TRS is preserved, but becomes activated when TRS is broken.} 
\end{table}

$\chi_{\rm I} (\omega)$ represents the first new result of our work and arises from the contrasting interlayer polarization when an electron transitions from state $n, \vec k \rightarrow m, \vec k$: its polarization changes from $p_{nn} \rightarrow p_{mm}$. 
As we will argue below, this process also yields an anomalous photoinduced interlayer current, controlled by an interlayer conductivity $\sigma_{\rm inter}^{\alpha\beta}(\omega)$. This anomalous interlayer current acts as a source that pumps the interlayer electric polarization. As a result, $\chi_{\rm I} (\omega)$ grows with $\tau$ yielding large LPE. This picture is similar to how bulk injection photocurrents are often understood as arising from a photoinduced acceleration~\cite{PhysRevX.10.041041,deJuan2017}.

Injection LPE contrasts with that of the shift LPE, $\chi_{\rm S}(\omega)$, recently discussed in Ref.~\cite{PhysRevLett.124.077401}: 
 \begin{equation}
    \chi_{\rm{S}}^{\alpha\beta}(\omega)=\pi  \frac{e^2}{\hbar^2} 
    \sum_{n,m, {\vec k}} \delta(\omega+\omega_{nm})f_{nm}
    A^\alpha_{nm} \mathcal{M}_{mn}^\beta,
    \label{eq:shift}
\end{equation}
where 
\begin{equation}
    \mathcal{M}_{mn}^\beta = \partial^\beta \frac{p_{mn}}{\omega_{mn}}-i{\sum_{c}}\left[A^\beta_{mc}\frac{\bar p^z_{cn}}{\omega_{cn}}-\frac{\bar p^z_{mc}}{\omega_{mc}}A^\beta_{cn}\right] ,
\end{equation}
and $\bar p^z_{nm}=p^z_{nm}(1-\delta_{nm})$. 
$\chi_{\rm S}$ is intrinsic ($\tau$ independent) and arises from
an interlayer coordinate shift that is non-vanishing in chiral media. 
 
In contrast to the other interband responses, $\chi_{\rm FS} (\omega)$ 
does not require real transitions. Instead, it corresponds to nonlinear {interlayer polarization} sustained even for light with frequency below the bandgap of an insulator. It is written as
$\chi_{\rm{FS}}^{\alpha\beta}(\omega)=\chi_{\rm{FS,1}}^{\alpha\beta}(\omega)+\chi_{\rm{FS,2}}^{\alpha\beta}(\omega)$,
where
\begin{equation}
    \chi_{\rm{FS,1}}^{\alpha\beta}(\omega)={\frac{ e^2}{2\hbar^2}}
    \sum_{n,m, {\vec k}}A^\alpha_{nm}A^\beta_{mn}f_{nm}\mathcal{P}\left[\frac{p^z_{mm}-p^z_{nn}}{(\omega+\omega_{nm})^2}\right],
    \label{eq:FS1}
\end{equation}
\begin{equation}
    \chi_{\rm{FS,2}}^{\alpha\beta}(\omega)=
    i  \frac{e^2}{\hbar^2}\sum_{n,m, {\vec k}}f_{mn}A^\alpha_{nm}\mathcal{M}_{mn}^\beta \mathcal{P}\left[\frac{1}{\omega+\omega_{nm}}\right],
    \label{eq:FS2}
\end{equation}
where $\mathcal{P}$ denotes the principal part. Strikingly, this off-resonant LPE survives even for insulators (unlike its photocurrent counterpart \cite{PhysRevResearch.3.L042032,PhysRevX.11.011001}). As a result, we denote it Fermi Sea LPE since it arises from virtual processes between completely occupied and unoccupied bands. $\chi_{\rm FS}$ proceeds in much the same fashion as that of the conventional dielectric response in insulators where similar virtual processes contribute to the dynamical screening. Indeed, $\chi_{\rm FS}$ can be understood as its nonlinear rectified counterpart. 

The last new LPEs we unveil are intraband in nature: these depend on the presence of a Fermi surface and exhibit a low-frequency divergence characteristic of metallic responses in the clean limit. 
These are the semiclassical (SC) and Berry (B) LPE responses: $ \chi_{\rm{intra}}(\omega) = \chi_{\rm{SC}}(\omega)+\chi_{\rm{B}}(\omega)$, with SC susceptibility: 
\begin{equation}
     \chi_{\rm{SC}}^{\alpha\beta}(\omega)= \frac{ e^2}{2\hbar ^2} 
     \sum_{n, {\vec k}}\frac{\partial^\alpha \partial^\beta f_n}{\omega^2+\tau^{-2}}p^z_{nn}, 
     \label{eq:sc}
\end{equation}
and Berry susceptibility:
\begin{equation}
     \chi_{\rm{B}}^{\alpha\beta}(\omega)=  \frac{e^2}{\hbar^2}
     \sum_{n,m, {\vec k}}\frac{p^z_{nm}A^\alpha_{mn}i\partial^\beta f_{nm}}{(\omega+i\tau^{-1})\omega_{nm}},
     \label{eq:m}
\end{equation}
where intra-band responses are regularized with a relaxation time $\tau$ \cite{PhysRevLett.123.246602}.
Note that $\chi_{\rm B}$ shares a similar density matrix origin to its counterpart in the more familiar but distinctly different photocurrent response (the Berry curvature dipole induced nonlinear Hall effect~\cite{PhysRevLett.115.216806}).  

$\chi_{\rm SC}(\omega)$ has a semiclassical origin: it arises from a DC shift (in momentum space) of the metallic Fermi surface induced by periodic driving; this enables to pick out a dipolar distribution of $p_{nn} (\vec k)$ in momentum space. $\chi_{\rm{B}}(\omega)$ arises from interband coherences sustained from the periodic driving; unlike the other responses we have discussed, $\chi_{\rm{B}}(\omega)$ has an odd-parity under time-reversal (c.f. $\partial^\beta f$ term), vanishing in non-magnetic materials. In what follows, we will focus on LPEs in time-reversal symmetry (TRS) preserving systems. 

\textit{\color{blue}Intrinsic out-of-plane interlayer current}. We now proceed to argue that the origin of the large injection LPE arises from an anomalous 
{\it out-of-plane} interlayer current induced by oscillating {\it in-plane} electric fields. To see this, we note that the interlayer electric current is naturally described by $\hat{j}^z ={d \hat P_z}/{dt} = [\hat P_z,\hat H]/(i\hbar)$ \cite{Resta2007}. Computing the expectation value of the interlayer current, $\langle j^z (t) \rangle$, we find
\begin{equation}
\label{eq:interlayercurrent}
{\rm Tr} [\hat{j}^z \rho(t)]  = \frac{1}{i\hbar}{\rm Tr} \left\{[\hat P_z,\hat H] \rho(t)\right\}  
=  {\rm Tr}[P^z \dot \rho(t)],
\end{equation}
where we have noted the cyclic property of the trace ${\rm Tr}\left\{[A,B]C\right\} = {\rm Tr}\left\{A[B,C]\right\}$ as well as employed the Liouville equation $i\hbar d\hat{\rho}(t)/dt = [\hat{H}(\mathbf{k},t),\hat{\rho}(t)]$. In order to isolate the rectified interlayer current, we focus on the period average $j^z_{\rm rectified} = [\int_0^T dt \, {\rm lim}_{\eta \to 0} \, \langle j^z(t) \rangle]/T$ where $T = 2\pi/\Omega$ is the period of the drive EM field. For a finite drive frequency $\Omega$, this directly produces an out-of-plane interlayer current 
\begin{equation}
    j^{z}_{\rm rectified} =2d{\rm Re}\left[E^{\alpha } E^{\beta*}\sigma^{\alpha\beta}_{\rm inter}(\Omega)\right],
\end{equation}
that is driven by an oscillating in-plane electric field $\vec E$. Here $\sigma^{\alpha\beta}_{\rm inter}(\Omega)$ is the interlayer nonlinear conductivity found in Eq.~(\ref{eq:inj}). Interestingly, Eq.~(\ref{eq:inj}) depends only on intrinsic band geometric quantities (e.g., $A_{nm}^\alpha$, $\delta P_{mn}$).  

We note that second-order nonlinear photocurrent susceptibilities have recently been the subject of intense investigation \cite{BerryDemon,2023arXiv230100811M,PhysRevB.23.5590,Sturman1992,PhysRevB.52.14636,PhysRevB.61.5337,BrehmYoung,Morimotoe1501524,NaMo,PhysRevB.99.045121,PhysRevLett.123.246602,PhysRevLett.127.126604,PhysRevB.19.1548,matsyshyn2020berry}. These have concentrated on photocurrents formed from bulk itinerant electrons with a well-defined velocity. In contrast, $\sigma^{\alpha\beta}_{\rm inter}(\Omega)$ describes out-of-plane current in a vdW stack hosting electrons that do not have a $z$-direction velocity. Instead, the interlayer current can be understood as a type of electric polarization pump that injects polarization.

\begin{figure*}
\centering
\includegraphics[width=\textwidth]{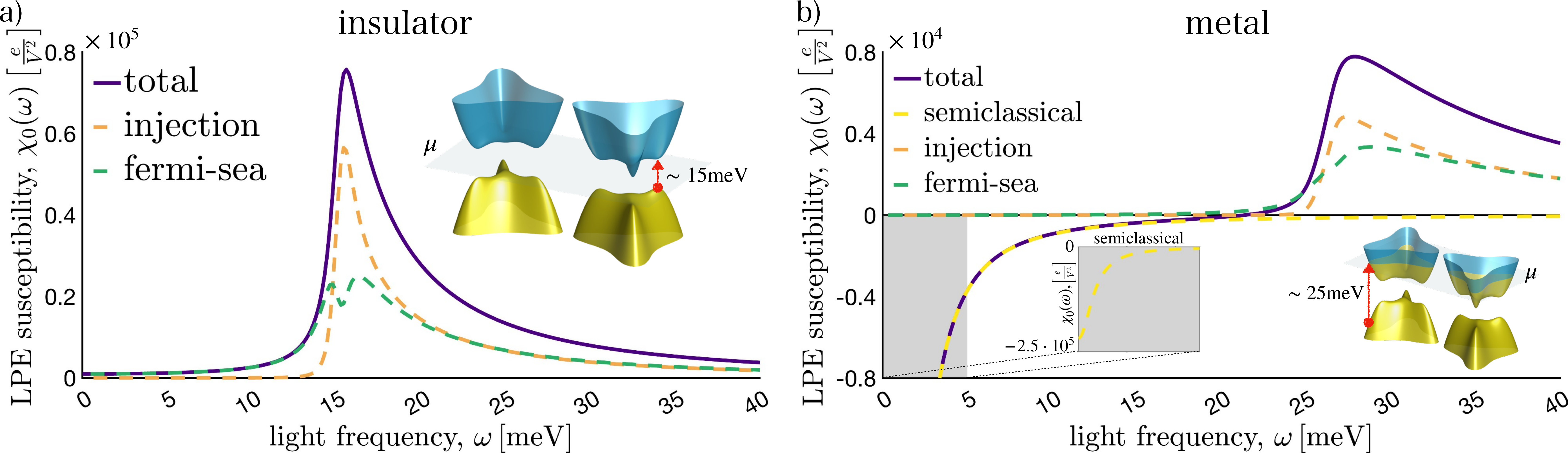}
\caption{{\it Nonhelical LPE responses in vdW heterostructure.} BLG/hBN LPE susceptibility tensor $\boldsymbol{\chi} (\omega)=\chi_0 (\omega) {\mathbb{I}}$ in the (a) insulating state ($\mu =10$ meV in the gap) (b) metallic state ($\mu =20$ meV) numerically evaluated using the low-energy Hamiltonian in Eq.~(\ref{GhBNHam}).  Both $\chi_{\rm I}$ (orange) and $\chi_{\rm FS}$ (green) contribute to the total response (purple) in the insulating state. In the metallic state, an additional metallic $\chi_{\rm SC}$ (yellow) emerges that dominates at low frequencies. Right inset in both panels display the low-energy bandstructure of BLG/hBN; $\mu$ indicates the Fermi level. Left inset in panel b shows a zoom-in of the gray region. Parameters used: $\tau =1\, {\rm ps}$ and $\Delta = 30\, {\rm meV}$.} 
\label{fig2}
\end{figure*}

\textit{\color{blue}Symmetry properties of LPE}. 
The mechanisms for LPE 
discussed above have distinct symmetry properties.
To see this, we re-write 
Eq.~(\ref{eq:LPEfreq}) as: 
\begin{multline}\label{polresL}
     \langle \delta{P}^z_{\rm st}\rangle/d = 
    \underbrace{\left(E^\alpha E^{*\beta}+E^{*\alpha} E^{\beta}\right)}_{\text{Linearly polarised light}}\underbrace{\frac12\left[\chi^{\alpha\beta}(\Omega)+\chi^{\alpha\beta}(-\Omega)\right]}_{\rm Re}+\\+\underbrace{\left(iE^\alpha E^{*\beta}-iE^{*\alpha} E^{\beta}\right)}_{\text{Circularly polarised light}}\underbrace{\frac{1}{2i}\left[\chi^{\alpha\beta}(\Omega)-\chi^{\alpha\beta}(-\Omega)\right]}_{\rm Im},
\end{multline}
displaying how the real (imaginary) parts of the susceptibility tensor control the response to linearly polarized (circularly polarized) irradiation. Recalling that under time-reversal symmetry we have $\vec A_{nm} (\vec k) = \vec A_{mn} (-\vec k)$ and $p_{nm} (\vec k) = p_{mn}(-\vec k)$ we obtain the non-helical (linear) vs helical (circular) classification in Table I, namely: $\chi_{\rm I}$, $\chi_{\rm FS}$ and $\chi_{\rm SC}$ mediate responses to linearly polarized light but are helicity insensitive; $\chi_{\rm S}$, in contrast, only arises under circularly polarized irradiation. Naturally, inversion symmetry zeroes out all LPE responses, see {\bf SI}. 

Point group symmetries also play a critical role in constraining the LPE. 
For instance, in-plane mirror symmetry $\mathcal{M}_y$ forces the off-diagonal components of the nonlinear LPE susceptibility tensor to vanish: $\chi^{xy}(\omega) = \chi^{yx}(\omega) =0$. This disables helicity dependent LPE. As a result, achiral vdW stacks (i.e. ones with a mirror plane) do not possess a helicity dependent LPE.
As a result, comparing with Table I, in these systems we find that LPE proceed from $\chi_{\rm I}$, $\chi_{\rm FS}$ and $\chi_{\rm SC}$ only; $\chi_{\rm S}$ vanishes.

In contrast, in chiral stacks that possess high crystalline symmetries, the opposite can be true. The combination of $C_{nz} ~(n\geq3)$ and $C_{2x}$ point-group rotational symmetries can render non-helical LPEs vanishing (see Ref.~\cite{PhysRevLett.124.077401} for an explicit example in twisted bilayer graphene as well as full symmetry analysis in {\bf SI}).  Of course, in chiral vdW stacks where at least one of these point group rotational symmetries are broken, both helicity dependent and non-helical LPEs are allowed. 

\textit{\color{blue} Non-helical LPE response in BLG/hBN}. To exemplify the non-helical LPE response from $\chi_{\rm I}$, $\chi_{\rm FS}$ and $\chi_{\rm SC}$ in TRS preserving 
systems, we focus on an achiral vdW system: bilayer graphene aligned with hexagonal Boron nitride (BLG/hBN). Aligned BLG/hBN breaks inversion symmetry, possesses $C_{3z}$ and $\mathcal{M}_y$ symmetries while breaking $C_{2x}$ (see Fig. 1). 
As a result, only non-helical LPE responses are allowed; $\chi_{\rm S}$ vanishes. Indeed, the presence of both $C_{3z}$ and $\mathcal{M}_y$ guarantee 
$\boldsymbol{\chi} (\omega)=\chi_0 (\omega) {\mathbb{I}}$, allowing LPE to manifest for unpolarized light.

We model the long-wavelength electronic excitations of BLG/hBN  using an minimal low-energy Hamiltonian 
\begin{equation}\label{GhBNHam}
    \hat H = \left(\begin{array}{cccc}
         \Delta/2&v\pi^\dagger &0 &v_3\pi \\
         v\pi&-\Delta/2 &\gamma_1 &0 \\
         0&\gamma_1 &0 &v\pi^\dagger \\
         v_3\pi^\dagger&0 &v\pi & 0
    \end{array}\right),
\end{equation}
where $v = 0.639\, \rm eV\, nm$ is the Dirac velocity of graphene, $v_3 = 0.081\,\rm eV\, nm$ characterizes trigonal warping, $\gamma_1 = 0.4\, \rm eV$ is the interlayer hopping, and $\pi = \xi k_x+ik_y$ where $\xi = \pm 1$ is the valley index. {Using Eq.~(\ref{eq:LPEoperator}) the polarization operator reads as} $\hat p_z = \rm diag(1,1,-1,-1)$. Responses of different valleys are added. $\Delta$ is the AB sublattice asymmetry induced by aligning one side with hBN thereby breaking inversion symmetry and opening a gap in the spectrum (see inset in Fig.~\ref{fig2}). In what follows we will concentrate on low frequencies 
up to the terahertz range where large LPEs manifest. This is smaller than 
the energy range ($150-200 \, {\rm meV}$) where superlattice effects from the hBN alignment ensue~\cite{yankowitz2012emergence}. 

The LPE in BLG/hBN was numerically evaluated using Eqs.~(\ref{eq:inj}), (\ref{eq:FS1}), (\ref{eq:FS2}),~and~(\ref{eq:sc}) at low temperature and summed across both valleys for the electronic states in Eq.~(\ref{GhBNHam}); LPE susceptibilities are plotted in Fig.~\ref{fig2}, see {\bf SI} for a full discussion of the numerical details. 
We find interband LPEs 
peak for frequencies close to the gap size, see Fig.~\ref{fig2}a where $\chi_{\rm I}$ and $\chi_{\rm FS}$ are plotted when the chemical potential is in the gap. This indicates that both $\chi_{\rm I}$ (orange) and $\chi_{\rm FS}$ (green) are dominated by interband processes close to the band edge. 

Interestingly, when the chemical potential is moved into the conduction band (Fig.~\ref{fig2}b), a new metallic peak in the nonlinear LPE response emerges at low frequencies that corresponds to $\chi_{\rm SC}$ (yellow); the interband LPE responses still persist but now appear at higher frequencies due to Pauli blocking (see right inset). The metallic peak is particularly striking since it displays large responses (left inset) even for frequencies below any interband optical transition, as well as the opposite sign of susceptibilities as compared to the interband contributions.  

The LPE we unveil demonstrates how stacking can introduce new classes of responses not found in a single layer. Indeed, we anticipate that $\chi_{\rm I}$ and $\chi_{\rm SC}$ can produce large LPE several orders of magnitude larger than that previously known, e.g., in Ref.~\cite{PhysRevLett.124.077401}. For instance, close to the interband peak in BLG/hBN heterostructures, we find a large interlayer surface charge density difference of order $1 ~{\rm n C} \, {\rm cm}^{-2}$ (this corresponds to an interlayer voltage of order $2 \, {\rm mV}$) can be sustained even for modest light intensity of $ 1000 \, {\rm W} \, {\rm cm}^{-2}$. At very low frequencies, LPE is expected to be even more pronounced, yielding up to $5\, {\rm mV}$ interlayer voltage under the same light intensity (see Fig.\ref{fig2}b left inset). Such interlayer voltages can be readily detected using capacitive probes \cite{PhysRevB.84.085441,PhysRevB.85.235458} or scanning electron transistors \cite{PhysRevLett.105.256806}, and are not just confined to BLG/hBN (that we have focussed on for a concrete illustration). Indeed, we expect that LPEs are generic and will manifest in the wide zoo of noncentrosymmetric layered heterostructures available, e.g., layered transition metal dichalcogenides. In addition to providing novel means of photodetection (especially in the THz regime), given the large LPE susceptibilities, the photoinduced interlayer polarizations may even enable light-driven means of switching the electric polarization in a range of vdWs layered ferroelectrics that have recently become available \cite{Zheng2020,Wang2022,yasuda2021stacking}. 

\textit{\color{blue}Acknowledgements}. This work was supported by the Ministry of Education Singapore under its MOE AcRF Tier 3 Grant No.~MOE~2018-T3-1-002 and a Nanyang Technological University start-up grant (NTU-SUG).

\bibliography{mysuperbib}

\newpage

\renewcommand{\theequation}{S\arabic{equation}}
\renewcommand{\thefigure}{S\arabic{figure}}
\renewcommand{\thetable}{S\Roman{table}}
\makeatletter
\makeatother
\setcounter{equation}{0}
\setcounter{figure}{0}
\setcounter{table}{0}

\section{Supplementary Information for ``Layer photovoltaic effect in van der Waals heterostructures''}

\subsection{Density matrix and perturbation theory}
In this section, we discuss perturbative corrections to the density matrix in the presence of an irradiating electromagnetic field; this is used to directly compute the interlayer polarization responses found in the main text.
Starting from the Liouville equation:
$i\hbar d\hat{\rho}(t)/dt= [\hat{H}(\mathbf{k},t),\hat{\rho}(t)]$,
with electric field employed in the length gauge $\hat{H}(\mathbf{k},t) = H_{0}(\mathbf{k})+e\hat{\mathbf{r}}\cdot\mathbf{E}(t)e^{\eta t}$, we compute perturbative corrections for density matrix (DM):
$\hat{\rho} =\hat{\rho}^{(0)}+\hat{\rho}^{(1)}+\hat{\rho}^{(2)}+\mathcal{O}(\mathbf{E}^3),$ where the index (0,1,2) represents the order of corrections. The second order correction is given by:
\begin{multline}\label{SIdm2}
    \rho^{(2)}_{nm}(t)=\frac{e^2}{\hbar ^2}\iint \frac{d\omega_2d\omega_1}{(2\pi)^2}E^{\alpha}(\omega_2)E^{\beta}(\omega_1)e^{-i(\omega_1+\omega_2)t+2\eta t}\\\times\Bigg\{
    \delta_{nm}\frac{i\partial^\beta i\partial^\alpha f_n}{(\omega_1+\omega_2+2i\eta)(\omega_2+i\eta)}+\\+\frac{1}{\omega_1+\omega_2-\omega_{nm}+2i\eta}i\partial^{\beta}\frac{A^{\alpha}_{nm}f_{mn}}{\omega_2-\omega_{nm}+i\eta}+\\+\frac{A^{\beta}_{nm}i\partial^\alpha f_{mn}}{(\omega_{2}+i\eta)(\omega_1+\omega_2-\omega_{nm}+2i\eta)}+\\\frac{1}{\omega_1+\omega_2-\omega_{nm}+2i\eta}\sum_c\left[\frac{A^\beta_{nc}A^{\alpha}_{cm}f_{mc}}{\omega_2-\omega_{cm}+i\eta}-\frac{A^\alpha_{nc}A^{\beta}_{cm}f_{cn}}{\omega_2-\omega_{nc}+i\eta}\right]\Bigg\}.
\end{multline}

Using Eq.~(\ref{SIdm2}) and Eq.~(\ref{eq:LPEoperator}) in the main text, the total polarization response is given by:
\begin{equation}
\label{eq:SIpolarization}
    \langle P^z\rangle /d= \sum_{i=0}^\infty\int\frac{d\mathbf{k}}{(2\pi)^2}\sum_{nm} p^z_{nm}\rho^{(i)}_{mn}.
\end{equation}
In the main text, we focussed on the 
$i = 2$ contribution since it gives the leading DC (i.e. static) {\it photoinduced} interlayer polarization. For monochromatic light, the DC contribution can be obtained after averaging the polarization over one period: $\langle f_{\rm st}\rangle = \int_0^T f(t)dt/T$. In so doing we concentrated on $\omega_1+\omega_2 =0$ contributions in Eq.~(\ref{SIdm2}). 

The injection [Eq.(\ref{eq:inj})], shift [Eq.(\ref{eq:shift})], and Fermi-Sea [Eqs.(\ref{eq:FS1}) and (\ref{eq:FS2})] contributions to the LPE in the main text can be obtained by plugging the terms in the third and fifth lines of Eq.(\ref{SIdm2}) into Eq.~(\ref{eq:SIpolarization}). 
The contributions can be naturally delineated into resonant (includes delta functions, e.g., for $\chi_{\rm I}$ and $\chi_{\rm S}$) and off-resonant (involves principal parts, e.g., $\chi_{\rm FS}$) in the limit of vanishingly small $\eta \to 0$. Similarly, the SC and Berry contributions to the LPE can be obtained directly by substituting the terms in the second and fourth lines of Eq.(\ref{SIdm2}) into Eq.~(\ref{eq:SIpolarization}) respectively.

\subsection{Interlayer current}
In this section, we provide a fuller account of the intrinsic interlayer current. First, for the convenience of the reader, we recall that the interlayer current operator can be written as:
\begin{equation}
    \hat{j}^z = \frac{d \hat P^z}{dt} = \frac{1}{i\hbar}[\hat P^z, \hat H].   
\end{equation}
As a result, the time varying interlayer current can be directly evaluated as 
\begin{multline}
    \langle j^{z}(t)\rangle ={\rm Tr}[\hat j^z \hat \rho(t)]=\frac{1}{i\hbar}{\rm Tr}\left\{[\hat P^z, \hat H]\hat\rho(t)\right\}=\\= \frac{1}{i\hbar}{\rm Tr}\left\{\hat P^z [\hat H,\hat\rho(t)]\right\},
\end{multline}
where we used the cyclic permutation property of the trace:
${\rm Tr}\left\{[A,B]C\right\} = {\rm Tr}\left\{A[B,C]\right\}.$

Importantly, the Liouville equation for the full system requires 
$i\hbar{d}\hat{\rho}(t)/{dt}= [\hat{H},\hat{\rho}(t)]$. As a result, we can directly identify the interlayer current as
\begin{equation}
     \langle j^{z}(t) \rangle ={\rm Tr}\left\{\hat P^z {{\dot{\rho}}(t)}\right\},
\end{equation}
thus reproducing Eq.~(\ref{eq:interlayercurrent}) of the main text. The rectified component of the interlayer current can be directly obtained as the period average of $\langle j^{z}(t) \rangle $: 
\begin{equation}
    j^{z}_{\rm rectified} =\int_0^T\frac{dt}{T}\lim_{\eta\rightarrow 0}{\rm Tr}\left\{\hat P^z {{\dot{\rho}_n}(t)}\right\}+\mathcal{O}(E^4).
    \label{eq:periodaverage}
\end{equation}
Crucially, it is the time derivative of the density matrix that controls the rectified interlayer current response. By directly taking a time derivative of the second order correction to the density matrix in Eq.~(\ref{SIdm2}), we find that there are only two terms in Eq.~(\ref{SIdm2}) that generate a finite contribution to the current above: i.e. terms in $\rho_{nm}^{(2)} (t)$ that originally corresponded to the injection and semiclassical LPEs. Notice, however, that the latter contribution to the intrinsic interlayer current in Eq.~(\ref{eq:periodaverage}) [after contraction with electric fields and symmeterizing $\alpha\leftrightarrow \beta, \omega\leftrightarrow -\omega$] displays a delta function peak at zero frequency $\delta (\omega)$. As a result, at finite non-zero frequencies only the injection type response remains in the limit $\eta \to 0$. 

\subsection{Time-reversal, inversion and spatial symmetries}
This section will briefly dicuss constraints dictated by time-reversal, inversion and spatial symmetries for the LPE susceptibility tensor. 

We first focus on time-reversal symmetry ($\mathcal{T}$). TRS produces the following relations for the matrix elements of the Berry connection,
polarization operator, Berry connections, as well as the band energy difference: $\mathbf{A}_{nm}(\mathbf{k})=\mathbf{A}_{mn}(-\mathbf{k})$, $\omega_{nm}(\mathbf{k})=\omega_{nm}(-\mathbf{k})$, $p_{nm}(\mathbf{k}) = p_{mn}(-\mathbf{k})$. Applying these relationships, we find that the Berry LPE response [in Eq.~(\ref{eq:m})] vanishes in the presence of TRS (since $\partial_k f$ is odd while $\vec A_{\rm nm}$ is even under $\vec k \to - \vec k$). In contrast, in the presence of TRS, SC LPE response persists. The interband LPE responses, however, have susceptibilities that obey 
\begin{gather}
    \chi^{\alpha\beta}_{\rm FS}(\omega)=\chi^{\alpha\beta}_{\rm FS}(-\omega)=\chi^{\beta\alpha}_{\rm FS}(\omega),\\\chi^{\alpha\beta}_{\rm S}(\omega)=-\chi^{\alpha\beta}_{\rm S}(-\omega),\\
    \chi^{\alpha\beta}_{\rm I}(\omega)=\chi^{\alpha\beta}_{\rm I}(-\omega)=\chi^{\beta\alpha}_{\rm I}(\omega),
\end{gather}
in the presence of TRS. 

Note that in a similar way, the presence of inversion symmetry ($\mathcal{I}$) demands:
$\mathbf{A}_{nm}(\mathbf{k})=-\mathbf{A}_{nm}(-\mathbf{k})$, $\omega_{nm}(\mathbf{k})=\omega_{nm}(-\mathbf{k})$, $p_{nm}(\mathbf{k}) = -p_{nm}(-\mathbf{k})$. As a result, inversion symmetry forces all second-order LPE responses to vanish; broken inversion symmetry is required to realize LPE responses as expected. 

\begin{table}[t]
\centering
\resizebox{0.48\textwidth}{!}{
\begin{tabular}{|C{1.4cm}||C{1.2cm}|C{1.2cm}| C{1.2cm} | C{1.4cm}| }
\hline\hline&&&&\\
  & $\mathcal{M}_y$  &$C_{3z}$&$C_{2x}$&$C_{2x} + C_{3z}$ \\&&&&\\\hline\hline{}&{}&{}&{}&{}\\
 LPL,~\faArrowsV & \cmark  &{\cmark}&{\cmark}&{\xmark}\\
 {}&{}&{}&{}&{}\\\hline{}&{}&{}&{}&{}\\
 CPL,~\faRepeat & \xmark  &{\cmark}&\cmark&\cmark\\
 {}&{}&{}&{}&{}\\\hline
\end{tabular}}
\caption{\label{tab:Table2} This table summarises whether LPE responses are allowed (\cmark) or forbidden (\xmark) in systems exposed to incident light with either circular (CPL) or linearly (LPL) polarisation in the presence of spatial symmetries 
$C_{2x}, C_{3z}$ and $\mathcal{M}_y$.}
\label{tablepoint}
\end{table}

Point group symmetry play an additional critical role in constraining LPE responses. For instance, in-plane mirror symmetry $\mathcal{M}_y$ forces off-diagonal components of the susceptibility tensor to vanish $\chi^{xy}(\omega) = \chi^{yx}(\omega) =0$. As a result, in-plane mirror symmetry disables helicity dependent second-order interlayer polarization responses (see column 2 of Table \ref{tablepoint}). Indeed, as discussed in the main text, BLG/hBN possess in-plane mirror symmetry, zeroing the helicity dependent $\chi_{\rm S}$ response. 

We now discuss the impact of point group rotational symmetry. For instance, in the presence of $C_{nz} ~(n\geq3)$ symmetry, the LPE susceptibility tensor obeys $\chi^{xx}(\omega) = \chi^{yy}(\omega)$, $\chi^{xy}(\omega) = -\chi^{yx}(\omega)$. Similarly, in the presence of $C_{2x}$ symmetry, the out of plane polarization has to switch its sign $\langle P_z\rangle \rightarrow-\langle P_z\rangle$ {under the operation of $C_{2x}$}: this means that the susceptibility components that preserve their sign under $C_{2x}$ are forced to vanish $\chi^{xx}(\omega) = \chi^{yy}(\omega)=0$. However, off-diagonal components of the LPE susceptibility tensor are allowed. Note that the off-diagonal components of the LPE susceptibility tensor in principle encode both helicity dependent [i.e. anti-symmetric part: $\chi^{xy}(\omega)-\chi^{yx}(\omega)$] as well as nonhelical responses to linearly polarized light [i.e. symmetric part: $\chi^{xy}(\omega)+\chi^{yx}(\omega)$]. Crucially, the presence of just one of the above symmetries {\it alone}, is compatible with both helical as well as non-helical responses (see column 3 and 4 of Table ~\ref{tablepoint}). However, when both $C_{3z}$ and $C_{2x}$ symmetries are present simultaneously (e.g., in pristine twisted bilayer graphene), they ensure that only helicity dependent LPE responses manifest (in the case of TRS, only $\chi_{\rm S}$ is non-zero); non-helical responses in such systems with both $C_{3z}$ and $C_{2x}$ vanish.   

\subsection{Details of numerics}
The evaluation of LPE responses for BLG/hBN shown in the main text was carried out numerically by using the BLG/hBN Hamiltonian in Eq.~(\ref{GhBNHam}) of the main text as well as the expressions for the various LPE responses found in the main text. In so doing, we used an effective relaxation time $\tau = 1\, {\rm ps}$ as an illustration that is characteristic of ultraclean graphene based heterostructures~\cite{wang2013one}. Further, in numerically evaluating integrals with delta functions and principal values, we compared the LPE response expressions directly with the density matrix in Eq.~(\ref{SIdm2}), using their corresponding finite but small $\eta $ representations (and taking $\eta \to 1/\tau$). Moreover, we found our numerically evaluated LPE responses converged rapidly at low temperature.

While $\chi_{\rm I}$, $\chi_{\rm S}$ depend on the interband transition contours and $\chi_{\rm SC}$ depends on the Fermi surface, the Fermi sea responses $\chi_{\rm FS}$ in Eqs.~(\ref{eq:FS1}) and (\ref{eq:FS2}) of the main text, in principle, depend on an entire band sum. Here we evaluated $\chi_{\rm FS}$ systematically starting from the Dirac point and moving outwards in momentum space; importantly we achieved convergence rapidly for $v|\mathbf{k}_{\rm max}|\sim 100$ meV within the range of validity of our low-energy Hamiltonian. The grid for the momentum space integration was chosen to be $300 \times 300 $ k-points for the interband terms. For $\chi_{SC}$ we adopted a finer mesh with up to $900 \times 900$ k-points to capture the sharpness of the Fermi surface at low temperatures. 

\section{Interlayer polarization and interlayer voltage}
While the photoinduced interlayer polarizations can be directly obtained in Eq.~(\ref{eq:LPEfreq}) from the LPE susceptibilities and oscillating electric fields of the incident EM irradiation, these interlayer polarizations can also manifest as an {\it interlayer voltage}, $\Delta U$. Modeling the bilayer system as a simple parallel plate capacitor separated by an interlayer distance of $2d = 3.46~{\rm\AA}$, we find    
\begin{equation}
    \Delta U = \frac{2d}{\varepsilon_0}\frac{\delta\sigma}{2}=\frac{\langle \delta P^z_{\rm static}\rangle}{2\varepsilon_0},
\end{equation}
where we used $\varepsilon_0$ for the vacuum permittivity between the graphene planes, and $\delta\sigma$ (with units $[\rm C\cdot cm^{-2}]$) is the photoinduced difference between the surface charge densities on the top and bottom layers 
\begin{equation}
    \delta\sigma = \frac{\langle \delta P^z_{\rm static}\rangle}{2d} =\sum_{\alpha\beta}{\rm Re}\left[E^\alpha E^{\beta*}\chi^{\alpha\beta}(\Omega)\right] 
\end{equation}
obtained directly from Eq.~(\ref{eq:LPEfreq}) in the main text.

\end{document}